\documentclass[10pt, conference, letterpaper]{IEEEtran}

\makeatletter
\def\ps@headings{%
\def\@oddhead{\mbox{}\scriptsize\rightmark \hfil \thepage}%
\def\@evenhead{\scriptsize\thepage \hfil \leftmark\mbox{}}%
\def\@oddfoot{}
\def\@evenfoot{}}
\def\blfootnote{\xdef\@thefnmark{}\@footnotetext}
\makeatother

\usepackage{cite}
\usepackage{url}
\usepackage{balance}

\usepackage{graphicx}
\usepackage{algorithm}
\usepackage{algorithmic}
\usepackage{subfigure}
\usepackage{amssymb, amsmath,graphicx,charter, latexsym}
\usepackage{enumerate}

\newtheorem{lemma}{Lemma}

\newtheorem{theorem}{Theorem}

\begin{document}
\title{Predictive Scheduling for Virtual Reality}
\author{\IEEEauthorblockN{I-Hong Hou, Narges Zarnaghi Naghsh}
\IEEEauthorblockA{Department of ECE\\
Texas A\&M University\\
College Station, TX 77840, USA\\
Email: \{ihou, nzarnaghi\}@tamu.edu}
\and
\IEEEauthorblockN{Sibendu Paul, Y. Charlie Hu}
\IEEEauthorblockA{School of ECE\\
Purdue University\\
West Lafayette, IN 47907, USA\\
Email: \{paul90, ychu\}@purdue.edu}
\and
\IEEEauthorblockN{Atilla Eryilmaz}
\IEEEauthorblockA{Department of ECE\\
Ohio State University\\
Columbus, OH, 43210, USA\\
Email: eryilmaz.2@osu.edu}
 }
\maketitle

\begin{abstract}

A significant challenge for future virtual reality (VR) applications is to deliver high quality-of-experience, both in terms of video quality and responsiveness, over wireless networks with limited bandwidth. This paper proposes to address this challenge by leveraging the predictability of user movements in the virtual world. We consider a wireless system where an access point (AP) serves multiple VR users. We show that the VR application process consists of two distinctive phases, whereby during the first (proactive scheduling) phase the controller has uncertain predictions of the demand that will arrive at the second (deadline scheduling) phase. We then develop a predictive scheduling policy for the AP that jointly optimizes the scheduling decisions in both phases. 

In addition to our theoretical study, we demonstrate the usefulness of our policy by building a prototype system. We show that our policy can be implemented under Furion, a Unity-based VR gaming software, with minor modifications. Experimental results clearly show visible difference between our policy and the default one. We also conduct extensive simulation studies, which show that our policy not only outperforms others, but also maintains excellent performance even when the prediction of future user movements is not accurate.
\end{abstract}
\section{Introduction} \label{section:introduction}

Virtual Reality (VR) is an emerging technology that has demonstrated great potential for commercial success. In addition to consumer adoption for video and gaming, it is also expected that VR will attract growing enterprise adoption in the areas of marketing, product demonstration, and training \cite{CCSreport}. While it is widely projected that the VR market will keep growing \cite{CCSreport, MMreport, SDreport}, it is expected that there will be a shift toward VR devices that can provide high quality of experience without wired tethering. 

Providing high-quality and wire-free VR experiences poses significant technology challenges. Current commercial wire-free VR devices mostly rely on self-contained VR headsets that handle all computation tasks, such as image rendering, with on-board processors. These so-called ``standalone'' VR devices, including Oculus Rift and Sony PlayStation VR, are limited by the processing power of on-board processors, and hence cannot deliver high-quality VR. On the other hand, recent studies \cite{abari2017enabling, abari2016cutting, lai2019furion} have proposed using a server, such as a PC or a game console, to process computation tasks and to stream high-quality images to VR headsets through wireless communications. However, such proposals require very high wireless bandwidth to achieve highly responsive VR experience with low motion-to-photon latency. 

Noticing that users' mobility patterns in the virtual world are usually predictable, this paper aims to design a predictive scheduling policy that proactively delivers rendered images to VR users even before users move in the virtual world. To design such a policy, we first provide an analytical model that jointly captures the features of playback process in VR applications, video encoding techniques, user mobility patterns in the virtual world, and wireless capacity. We observe that there are two distinct phases between two image playbacks: a \emph{proactive scheduling phase} before the users' movements are recorded, and a \emph{deadline scheduling phase} after the movements are recorded. The goal of the predictive scheduling policy is then to find the jointly optimal scheduling decisions in both phases so as to maximize the perceived quality of experience (QoE) of VR users.

We formulate the problem of maximizing QoE of VR users as an optimization problem. We develop an offline iterative algorithm that optimally solves the problem. However, this offline algorithm requires the server to have the precise knowledge of user mobility patterns. To address the challenge of potentially noisy estimation of user mobility, we also propose a simple online scheduling policy using the insights derived from the offline algorithm.

We also conduct comprehensive ns-2 simulations to evaluate the performance of our online scheduling policy. Simulation results show that our policy clearly outperforms several other baseline policies. They also show that our policy is able to maintain excellent QoE even when there are considerable errors in the estimation of user mobility patterns, and that our policy converges very fast.

Finally, in order to demonstrate that our policy is readily implementable on existing VR systems, we build a prototype system under Furion \cite{lai2019furion}, a Unity-based system. While several software constraints in Unity are not fully compatible with our analytical model, we show that our online policy can still be implemented with minor modifications. Experimental results with one user and one server communicating over an off-the-shelf wireless access point show that our policy is able to deliver a much better experience than the default policy in Furion when the wireless bandwidth is limited.

The rest of the paper is organized as follows: Section \ref{section:model} describes an analytical model for studying predictive scheduling algorithms and formulates the optimization problem of maximizing QoE for VR users. Section \ref{section:offline} proposes an iterative offline algorithm that optimally solves the optimization problem. Section \ref{section:policy} develops an online scheduling policy. Section \ref{section:simulation} presents our simulation results. Section \ref{section:prototype} discusses implementing our online scheduling policy on existing VR systems and demonstrates experimental results. Section \ref{section:related} summarizes some relevant existing studies. Finally, Section \ref{section:conclusion} concludes the paper.
\section{System Model} \label{section:model}

We consider a wireless system where an access point (AP) streams real-time interactive VR  data to a number of wireless users. Each user has a headset that plays VR 360-degree panoramic images that are periodically updated  based on the user's location. For example, at 60 frame-per-second ($fps$), the headset plays one 360-degree panoramic image every 16 $ms$. We say that the time between two image playbacks is an \emph{interval}, and an interval consists of $M$ \emph{time slots}, where the AP can transmit one packet in each time slot. 

\begin{figure}
  \centering
  \includegraphics[width = 3.2in]{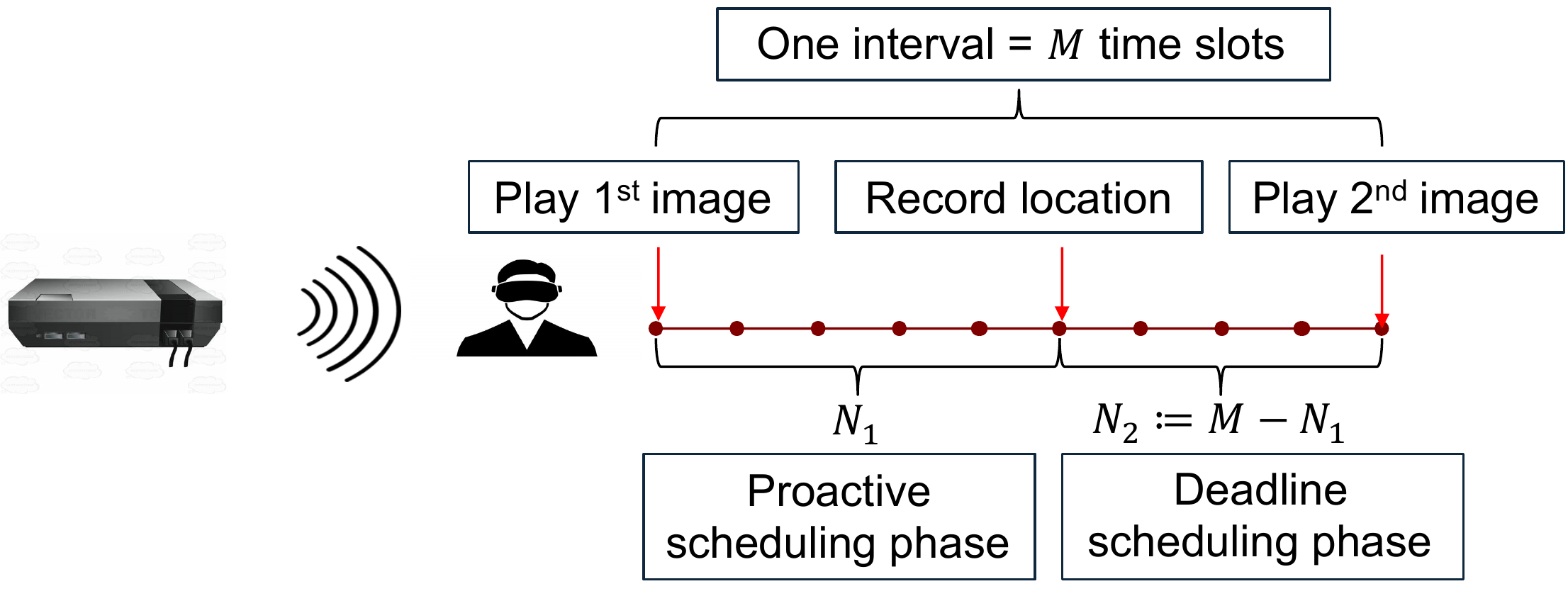}
  \caption{The timeline of one interval.}\label{figure:system:timeline}
\end{figure}

Users move in the virtual world following arbitrary mobility patterns. Each location in the virtual world corresponds to a different 360-degree panoramic image. After the first $N_1$ time slots in each interval, each headset records its user's latest movement, and sends the user's current location to the AP. After another $N_2:=M-N_1$ time slots, the headset displays the image corresponding to this current location. Fig. \ref{figure:system:timeline} shows the timeline of the system. We note that the value of $N_2$ loosely corresponds to motion-to-photon latency. Hence, when $M$ is fixed, smaller $N_2$ leads to more responsive VR experience.

We now discuss the scheduling policy of the AP. As shown in Fig. \ref{figure:system:timeline}, in the first $N_1$ time slots, the AP does not know the current locations of users for sure. However, it knows each user's location in previous intervals, and can use this information to estimate each user's current location. The AP can then transmit packets to users proactively based on its estimation. Hence, we call the first $N_1$ time slots in each interval as the \emph{proactive scheduling phase}. 
On the other hand, in the last $N_2$ time slots, the AP knows the exact current locations of users, and can make scheduling decisions accordingly, with the constraints that only images delivered before the end of the interval can be properly displayed. We hence call the last $N_2$ time slots in an interval as the \emph{deadline scheduling phase}. 

Due to the uncertainties in user mobility and the limited number of time slots in each interval, it is possible that the AP is unable to deliver all 360-degree panoramic images to users on time. We employ multi-layered video coding \cite{han2006method} to ensure smooth video playback even when some information is not delivered on time. By using multi-layered video coding, each 360-degree panoramic image is encoded into a base layer and several enhancement layers. A user is able to play a 360-degree panoramic image as long as it receives the base layer. The quality of experience (QoE) of the playback depends on the number of enhanced layers that a user receives. Obviously, the more are the number of received enhanced layers, the better the perceived QoE. Each layer is composed of a number of packets, and the value of a packet is determined by the impact of QoE on its corresponding layer. For example, a packet of the base layer has a higher value than that of an enhancement layer.

We assume that each user's mobility pattern can be described as a positive-recurrent irreducible Markov process over a finite state space $\mathcal{S}$. The state of the system in each interval is defined to be the collection of the state of each user, and we use $f_s$ to denote the steady-state probability that the system state is $s$.

When the state of the system in the previous interval is $s$, then the AP can estimate the probability that a user moves to a particular location, for each user. We say that a packet is \emph{wanted} if it corresponds to the current location of one of the users. We use $p_{s,i}$ to denote the probability that the packet $i$ will be wanted in the current interval when the system state in the previous interval is $s$. Different packets have different impacts on QoE, and we use $v_i$ to denote the value of the packet $i$.  

As described above, the proactive scheduling phase is the first $N_1$ time slots where the AP knows the state in the previous intervals, but not the state in the current interval. Let $x_{s,i}$ be the indicator function that the AP transmits the packet $i$ in the proactive scheduling phase. Obviously, we have $\sum_{i}x_{s,i} \leq N_1$, for all $s$, and $0 \leq x_{s,i}\leq 1 $ , for all $s$ and $i$.

On the other hand, during the deadline scheduling phase, the AP knows the current location of the users, and the packets that users want. Let $y_{s,i}$ be the probability of transmitting the packet $i$ in the deadline scheduling phase  when the system state in the previous interval is $s$. The AP can only transmit the packet $i$ in the deadline scheduling phase if the packet has not been transmitted already, i.e., $x_{s,i}=0$, and is wanted by users, which happens with probability $p_{s,i}$. Therefore, $y_{s,i} \leq p_{s,i}(1-x_{s,i})$. Also, as there are only $N_2$ time slots in the deadline scheduling phase, we have $\sum_{s}f_s\sum_{i}y_{s,i}\leq N_2$. We note that this inequality implicitly assumes that we only require the \emph{average} number of transmissions in the deadline scheduling phase to be no more than $N_2$. In Section \ref{section:policy}, we will derive an online policy which ensures that the number of transmissions in \emph{every} interval to be no more than $N_2$.

The QoE of a user depends on the packets that the user wants and have been delivered to him/her. If the packet $i$ is transmitted in the proactive scheduling phase, i.e. $x_{s,i}=1$, then it is wanted with probability $p_{s,i}$. Therefore, its expected contribution to QoE in the proactive scheduling phase is $x_{s,i}p_{s,i}v_i$. On the other hand, if the packet $i$ is transmitted in the deadline scheduling phase, which happens with probability $y_{s,i}$, then it is of course wanted by the users. Hence, its expected contribution to QoE in the deadline scheduling phase is $y_{s,i}v_i$. The total expected QoE contributed by packet $i$ under state $s$ is then $v_i (x_{s,i}p_{s,i}+y_{s,i})$.

Our goal is to maximize the total QoE, subject to all the aforementioned constraints, which can be written as the following linear programming problem:

    \begin{align}
\max\text{\quad} &\sum_{s}f_s\sum_{i}v_i \Big(x_{s,i}p_{s,i}+y_{s,i}\Big)
\label{1}\\
      \text{subject to \quad}  &\sum_{s}f_s\sum_{i}y_{s,i}\leq N_2, \label{2}\\
        &y_{s,i} \leq  p_{s,i}(1-x_{s,i}), \; \forall s, i,\label{3}\\
        &\sum_{i}x_{s,i} \leq N_1, \; \forall s, \label{4}\\
        &0\leq x_{s,i}\leq 1, \; \forall s, i, \label{5}\\
        & y_{s,i} \geq 0, \; \forall s, i.
        \label{7}
    \end{align}

Solving this linear programming problem requires the precise knowledge of $f_s$ and $p_{s,i}$. It can also be computationally infeasible. In the following sections, we first develop a low-complexity solution based on dual decomposition. We then show that this solution gives rise to a simple online scheduling algorithm that does not require the precise knowledge of $f_s$ and $p_{s,i}$.

\section{Iterative Offline Solution}\label{section:offline}

In this section, we develop an iterative offline solution by the dual decomposition method for solving the linear programming problem (\ref{1}) -- (\ref{7}) and prove that this solution is optimal. 

\subsection{Dual Decomposition Method}

We assign a Lagrange multiplier $\lambda$ to the constraint (\ref{2}). Let $\mathbb{X}$ and $\mathbb{Y}$ be vectors consisting all $x_{s,i}$ and all $y_{s,i}$, respectively. Then the Lagrangian $L(\mathbb{X},\mathbb{Y}, \lambda)$ is as follows:

\begin{align}
&L(\mathbb{X},\mathbb{Y}, \lambda) \nonumber\\
:=& \sum_{s}f_s\sum_{i}v_i \Big(x_{s,i}p_{s,i}+y_{s,i}\Big)-\lambda \Big(\sum_{s}f_s\sum_{i}y_{s,i}-N_2\Big)\nonumber\\
=& \sum_{s}f_s\Big[\sum_{i}v_i x_{s,i}p_{s,i}+\sum_{i}\Big(v_i-\lambda\Big)y_{s,i}\Big]+\lambda N_2. \label{8}
\end{align}

The dual objective function is as follows:

\begin{equation}
\begin{split}
D(\lambda):=&\max_{\mathbb{X}, \mathbb{Y}} L(\mathbb{X},\mathbb{Y}, \lambda),\\
& \text{subject  to} \quad (\ref{3})-(\ref{7}), \label{9}
\end{split}
\end{equation}



and the dual problem of (\ref{1}) -- (\ref{7}) is:

\begin{equation}
\begin{split}
\min_{\lambda \geq 0}D(\lambda). \label{10}
\end{split}
\end{equation}

It is easy to check that (\ref{1}) -- (\ref{7}) satisfy the Slater's condition, and hence, $\min_{\lambda \geq 0}D(\lambda)$ equals the maximum of the total QoE as mentioned in (\ref{1}) -- (\ref{7}). 

We note that the structure of (\ref{8}) and (\ref{9}) naturally provides a state-by-state decomposition. Specifically, it is easy to see that finding $D(\lambda)$ is equivalent to solving the following optimization problem for each state $s$: 

\begin{align}
\max\text{\quad}&\sum_{i}v_i x_{s,i}p_{s,i}+\sum_{i}\Big(v_i-\lambda\Big)y_{s,i} \label{11}\\
 \text{subject to}  \quad &y_{s,i} \leq  p_{s,i}(1-x_{s,i}), \;  \forall i,\label{12}\\
        &\sum_{i}x_{s,i} \leq N_1, \; \label{13}\\
        &0\leq x_{s,i}\leq 1, \; \forall i, \label{14}\\
        & y_{s,i} \geq 0, \; \forall i. \label{16}
\end{align}

Now suppose $x_{s,i}$ is given and fixed for each $s$ and $i$. Then, it is obvious that (\ref{11}) is increasing with $y_{s,i}$ if $\lambda \leq v_i$, and is decreasing with $y_{s,i}$ if $\lambda > v_i$. From (\ref{12}) and (\ref{16}), we have $0 \leq y_{s,i} \leq p_{s,i}(1-x_{s,i})$. Hence, the optimal choice of $y_{s,i}$, denoted by $y^*_{s,i}(\lambda)$ is 

\begin{equation}
\begin{split}
y^*_{s,i}(\lambda) = 
    \begin{cases}
      p_{s,i}(1-x_{s,i}), \quad & \text{if} \quad v_i \geq \lambda,\\
      0, \quad \quad \quad \quad & \text{if} \quad v_i < \lambda.
    \end{cases} \label{17}\\
\end{split}
\end{equation}

Substituting the above equation to (\ref{11}) -- (\ref{16}) yields:

\begin{equation}
\begin{split}
\max\text{\quad} &\sum_{i}v_i x_{s,i}p_{s,i}+\sum_{i}\Big(v_i-\lambda\Big)y_{s,i}\\
=&\sum_{i}v_i x_{s,i}p_{s,i}+\sum_{i:  v_i \geq \lambda} \Big(v_i-\lambda\Big)p_{s,i}\Big(1-x_{s,i}\Big)\\
=& \sum_{i} \alpha_{s,i}  x_{s,i}+ \sum_{i:v_i \geq \lambda} p_{s,i}\Big(v_i-\lambda\Big), \\
\text{s.t. \quad} &(\ref{13})-(\ref{14}), \label{18}
\end{split}
\end{equation}
where we define $\alpha_{s,i}$ as
\begin{equation}
\begin{split}
\alpha_{s,i} := 
    \begin{cases}
      v_i p_{s,i}, \quad \text{if} \quad v_i<\lambda,\\
      \lambda p_{s,i}, \quad \; \text{if} \quad v_i \geq \lambda. 
    \end{cases} \label{19}\\
\end{split}
\end{equation}

Let $x_{s,i}^*(\lambda)$ be the optimal solution for (\ref{18}). Obviously, the optimal solution is obtained by setting $x_{s,i}^*(\lambda)=1$ for the $N_1$ packets with the largest $\alpha_{s,i}$ , and $x_{s,i}^*(\lambda)=0$ for all of the other packets. Ties are broken by favoring packets with larger $p_{s,i}$. This together with $y_{s,i}^*(\lambda)$ defined in (\ref{17}) is the optimal solution to (\ref{9}) when $\lambda$ is given.


\subsection{Finding the optimal value of $\lambda$}

Next, we study the problem of finding the optimal $\lambda^*$ to minimize $D(\lambda)$. We follow a similar procedure proposed in \cite{Lin, Shor}. First, we find a subgradient for $D(\lambda)$.


\begin{lemma}
Let $\mathbb{X}^*(\lambda)=[x^*_{s,i}(\lambda)]$ and $\mathbb{Y}^*(\lambda)=[y^*_{s,i}(\lambda)]$ be the vectors that maximize $L(\mathbb{X},\mathbb{Y},\lambda)$, for a given $\lambda$. Then, $D'(\lambda) := N_2-\sum_{s}f_s\sum_{i}y^*_{s,i}(\lambda)$ is a subgradient of $D(\lambda)$. 
\end{lemma}

\begin{IEEEproof}
Let $\lambda'$ be an arbitrary value. We have:
\begin{align}
D(\lambda')&=\max_{x_{s,i},y_{s,i}}{L(\mathbb{X},\mathbb{Y},\lambda')}\nonumber\\
&\geq L(\mathbb{X}^*(\lambda),\mathbb{Y}^*(\lambda),\lambda')\nonumber\\
&=L(\mathbb{X}^*(\lambda),\mathbb{Y}^*(\lambda),\lambda)+(\lambda'-\lambda)D'(\lambda)\nonumber\\
&=D(\lambda)+(\lambda'-\lambda)D'(\lambda).
\end{align}
Thus, $D'(\lambda)$ is a subgradient of $D(\lambda)$.
\end{IEEEproof}

The following theorem is then a direct result following Theorem 8.9.2 in \cite{Baz}:

\begin{theorem} \label{theorem:gradient lambda}
Suppose ${h_t}$ is a sequence of non-negative step sizes such that $\sum_{t=0}^{\infty}h_t=\infty$, and $\lim_{t\rightarrow \infty} h_t =0$. Updating $\lambda (t)$ as follows:

\begin{equation}
    \begin{split}
&\lambda(t+1) = \bigg\{\lambda(t)-h_t\bigg[D'(\lambda(t))\bigg]\bigg\}^{+},
    \end{split}\label{21}
\end{equation}
ensures that, 
\begin{equation}
    \begin{split}
&\lim_{t \rightarrow \infty} D(\lambda (t)) = \min_{\lambda \geq 0} D(\lambda).
    \end{split}  \label{22}
\end{equation}

$\square$
\end{theorem}

Hence, by updating $\lambda$ iteratively according to (\ref{21}), we solve the dual problem. Alg. \ref{alg:offline} summarizes the complete algorithm of the iterative offline solution.

\begin{algorithm}
\caption{Offline Algorithm}\label{alg:offline}
\begin{algorithmic}[1]
\STATE Initialize $\lambda=0$
\FOR{$t=1, 2, \dots$} 
\FOR{each state $s$}
\FOR{each packet $i$} 
\STATE $x_{s,i}\gets 0$
\IF{$v_i \geq \lambda$}
\STATE $\alpha_{s,i} \gets p_{s,i} \lambda$
\ELSE
\STATE $\alpha_{s,i} \gets p_{s,i}v_i$
\ENDIF
\ENDFOR
\STATE Sort all packets so that $\alpha_{s,i_1}\geq \alpha_{s, i_2}\geq \dots$
\FOR{$j=1\to N_1$}
\STATE $x_{s,i_j}\gets 1$
\ENDFOR
\FOR{each packet $i$} 
\IF{$v_i \geq \lambda$}
\STATE $y_{s,i} \gets p_{s,i}(1-x_{s,i})$
\ELSE
\STATE $y_{s,i} \gets 0$
\ENDIF
\ENDFOR

\ENDFOR

\STATE $\lambda \gets \Big[\lambda-h_t[N_2-\sum_sf_s\sum_iy_{s,i}]\Big]^+$
\ENDFOR
\end{algorithmic}
\end{algorithm}

\section{Online Scheduling and Learning Policy} \label{section:policy}

While Section \ref{section:offline} provides an iterative algorithm that optimally solves (\ref{1}) -- (\ref{7}), the iterative algorithm cannot be directly turned into an implementable scheduling policy. First, the algorithm requires the precise knowledge of user mobility patterns, which is needed to calculate $f_s$ and $p_{s,i}$. Second, recall that the constraint $\sum_{s}f_s\sum_{i}y_{s,i}\leq N_2$ in (\ref{2}) only requires that the \emph{average} number of transmissions in the deadline scheduling phase to be no more than $N_2$. In practice, we need to ensure that the number of transmissions in \emph{every} interval to be no more than $N_2$.

In this section, we present a simple scheduling policy that addresses the aforementioned shortcomings. The policy consists of three parts: the policy in the proactive scheduling phase, the policy in the deadline scheduling phase, and the update rule of the Lagrange multiplier $\lambda(t)$, which is initialized by setting $\lambda(1)=0$.

The policy in the proactive scheduling phase is virtually the same as that in Section \ref{section:offline}. Let $s(t)$ be the system state in interval $t$, which contains the states of all individual users. During the proactive scheduling phase in interval $t$, the AP does not know $s(t)$ yet, but already knows $s(t-1)$. Given $s(t-1)$, the AP calculates $\alpha_i(t)$ as $\alpha_i(t)=v_i p_{s(t-1),i}$ if $v_i<\lambda(t)$, and $\alpha_i(t)=\lambda(t) p_{s(t-1),i}$ if $v_i\geq \lambda(t)$. The AP then transmits the $N_1$ packets with the largest values of $\alpha_i(t)$. As a result, a packet $i$ is transmitted in the proactive phase if and only if $x^*_{s(t-1),i}(\lambda(t))=1$.

Next, we discuss the policy in the deadline scheduling phase. After learning $s(t)$ in interval $t$, the AP sets $w_i(t)=1$ if the packet $i$ is wanted by users and has not been transmitted during the proactive scheduling phase, and sets $w_i(t)=0$, otherwise. The AP then sorts all packets by $v_i w_i(t)$ and transmits the $N_2$ packets with the largest $v_i w_i(t)$. Effectively, the AP transmits the $N_2$ packets with the largest values among those that are wanted by users and have not been transmitted yet. Given $s(t)$ and the policy for the proactive scheduling phase, this is indeed the optimal policy for the deadline scheduling phase that guarantees that the number of transmissions in the deadline scheduling phase is no more than $N_2$ in every interval.


Finally, we discuss the update of $\lambda(t)$. While Thm. \ref{theorem:gradient lambda} has established an iterative procedure for finding the optimal $\lambda$ for the dual problem, we note that calculating $D'(\lambda)=N_2-\sum_sf_s\sum_iy^*_{s,i}(\lambda)$ requires the precise knowledge about $f_s$, which is difficult to obtain in practice. Instead, we propose employing the stochastic gradient descent method for the update of $\lambda(t)$. After learning $s(t)$ in interval $t$, the AP sets $z_i(t)=1$ if $w_i(t)=1$ and $v_i\geq \lambda(t)$, and $z_i(t)=0$, otherwise. At the end of the interval $t$, the AP updates $\lambda(t)$ by $\lambda(t+1) = \Big[\lambda(t)-h_t\big(N_2-\sum_iz_i(t)\big)\Big]^+$.

We now show that this procedure finds the optimal $\lambda$ for the dual problem. Suppose $s(t-1)=s$, then $w_i(t)=1$ if and only if packet $i$ is not transmitted in the proactive scheduling phase, i.e., $x^*_{s,i}(\lambda(t))=0$, and is wanted by the users, which happens with probability $p_{s,i}$. Hence, we have
\begin{align}
&E[z_i(t)|s(t-1)=s]\nonumber\\ 
=& 
    \begin{cases}
      p_{s,i}(1-x_{s,i}^*(\lambda(t))),  & \text{if} \quad v_i \geq \lambda(t),\\
      0,  & \text{if} \quad v_i < \lambda(t),
    \end{cases}\nonumber\\
    =&y_{s,i}^*(\lambda(t)).
\end{align}
In steady state, we have $Prob(s(t-1)=s)=f_s$. Therefore, we have $E[N_2-\sum_iz_i(t)]=N_2-\sum_sf_s\sum_iy_{s,i}^*(\lambda(t))$. The following theorem is then a direct result from Theorem 46 in \cite{Sho} (section 2.4).

\begin{theorem} \label{theorem:sgd}
Suppose ${h_t}$ be a sequence of non-negative step sizes such that $\sum_{t=0}^{\infty}h_t=\infty$, $\sum_{t=0}^{\infty}h_t^2 \leq \infty$. If we update $\lambda(t)$ by $\lambda(t+1) = \Big[\lambda(t)-h_t\big(N_2-\sum_iz_i(t)\big)\Big]^+$, then $\lambda(t)$ converges to the optimal solution for the dual problem in probability. $\square$
\end{theorem}

Alg. \ref{alg:predictive} summarizes our online policy for predictive scheduling, where we simplify some of the notations to streamline the algorithm.

\begin{algorithm}
\caption{Predictive Scheduling for VR}\label{alg:predictive}
\begin{algorithmic}[1]
\STATE Initialize $\lambda=0$
\FOR{each interval $t$} 
\STATE \emph{// The proactive scheduling phase begins}
\FOR{each packet $i$} 
\IF{$v_i \geq \lambda$}
\STATE $\alpha_i \gets p_{s(t-1),i} \lambda$
\ELSE
\STATE $\alpha_i \gets p_{s(t-1),i}v_i$
\ENDIF
\ENDFOR

\STATE Sort all packets so that $\alpha_1 \geq \alpha_2 \geq ...$
\STATE Transmit packets $1 \sim N_1$
\STATE \emph{// The deadline scheduling phase begins}
\STATE {Obtain the locations of the users}
\STATE $w_i \gets 0, \forall i;$ $z_i \gets 0, \forall i$
\FOR{each wanted packet $i$ that was not sent}
\STATE $w_i \gets 1$
\IF{$v_i\geq\lambda$}
\STATE $z_i \gets 1$
\ENDIF
\ENDFOR
\STATE {Sort all packets so that} $v_1 w_1 \geq v_2 w_2 \geq ...$
\STATE {Transmit packets} $1 \sim N_2$
\STATE \emph{// Update the Lagrange multiplier $\lambda$}
\STATE $\lambda \gets \Big[\lambda-h_t[N_2-\sum_{i} z_{i}]\Big]^+$
\ENDFOR
\end{algorithmic}
\end{algorithm}

\section{Simulation Results} \label{section:simulation}

\subsection{Simulation Settings}
We have implemented our policy, as well as three other policies, in ns-2. We first discuss the settings of our simulation environment.

\textbf{Network environment:} Our setting for wireless transmissions follow the IEEE 802.11/ac standard. The system consists of one AP and several users. With 64-QAM and 80 MHz bandwidth, the AP can transmit at 1300 Mbps. Assuming that the maximum packet size is 2300 Bytes, ns-2 simulations show that the total time needed to transmit a packet, including all overheads such as the transmission of the ACK, is a bit smaller than 40 $\mu$s. After taking into account the amount of time needed for clients to transmit their locations to the AP, ns-2 simulations show that there are 399 slots in an interval of 16 ms, which corresponds to the scenario when the video playback rate is 60 fps.

\textbf{Video encoding:} We assume that multi-layer video coding is used and the panoramic frame of each location in VR environment is encoded into one base layer and four enhancement layers. We use publicly available video traces \cite{seeling2011video} to determine the parameters of each packet. In particular, the value of each layer is defined as the increment of frame quality over the previous layer. The value of a packet of a particular layer is then defined as the ratio of the value of the layer and the number of packets in that layer. Table \ref{tab:simulation:value} summarizes these parameters, where the first two columns are directly obtained from \cite{seeling2011video}, while the remaining columns are calculated based on the description above.

\begin{table}[h!]
\centering
\begin{tabular}{l|r|r|r|r}
Layer & Size (B) & Quality (dB) & \# of packets & value/packet    \\ \hline\hline
Base & 1615.96 & 27.924 & 1 & 27.924\\ \hline
Enh. 1  & 2048.22 & 32.012 & 1 & 4.088\\ \hline
Enh. 2  & 6761.27 &37.408 & 3 & 1.798 \\ \hline
Enh. 3  & 26060.27 & 42.446 & 12 & 0.419\\ \hline
Enh. 4  & 60609.8 & 47.708 & 27 & 0.194
\end{tabular}
\caption{Parameters for different layers} \label{tab:simulation:value}
\end{table}

\begin{figure*}[t]
\subfigure[10 clients]{
\includegraphics[width=2.1in]{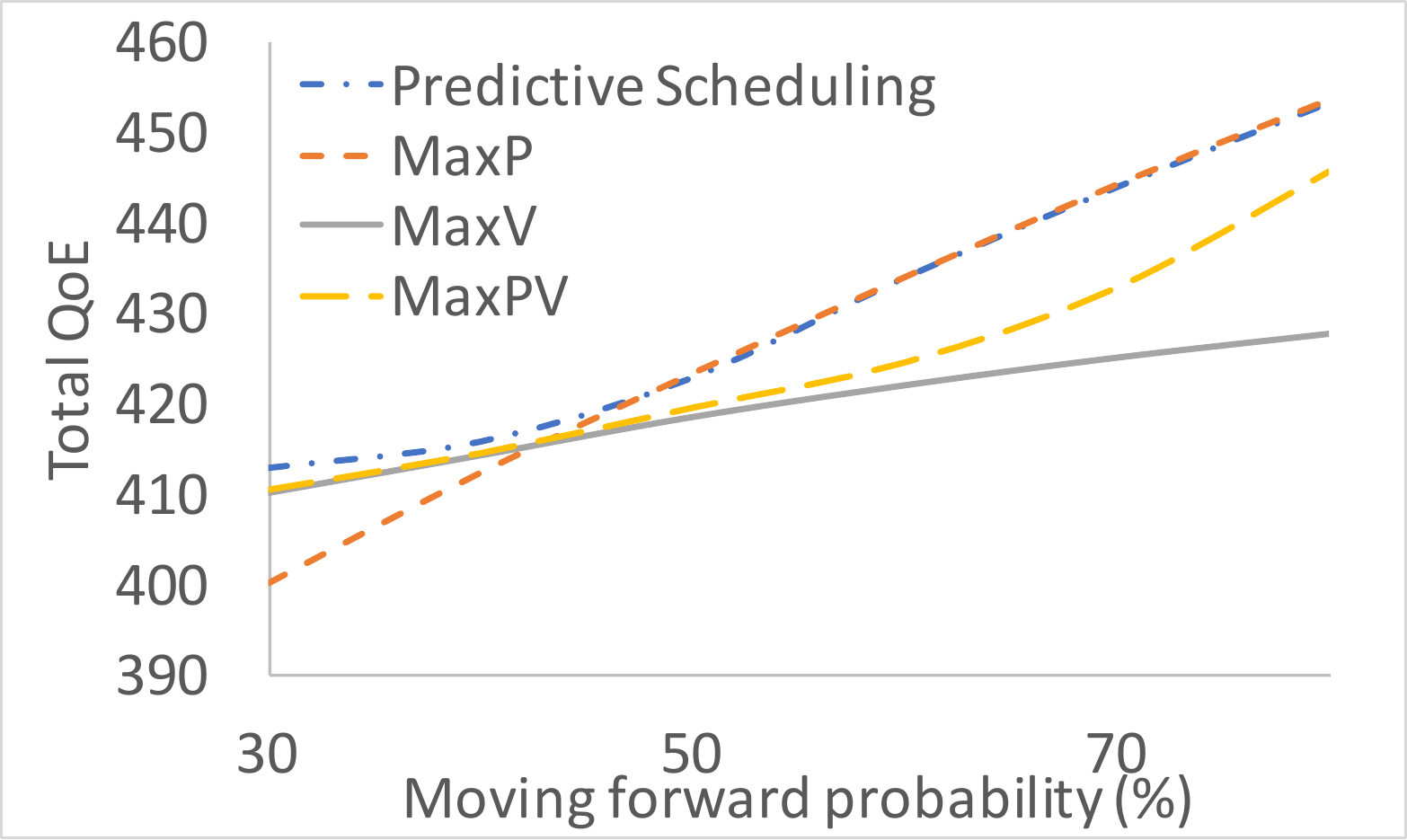}}
\hspace{0.01\linewidth} 
\subfigure[15 clients]{
\includegraphics[width=2.1in]{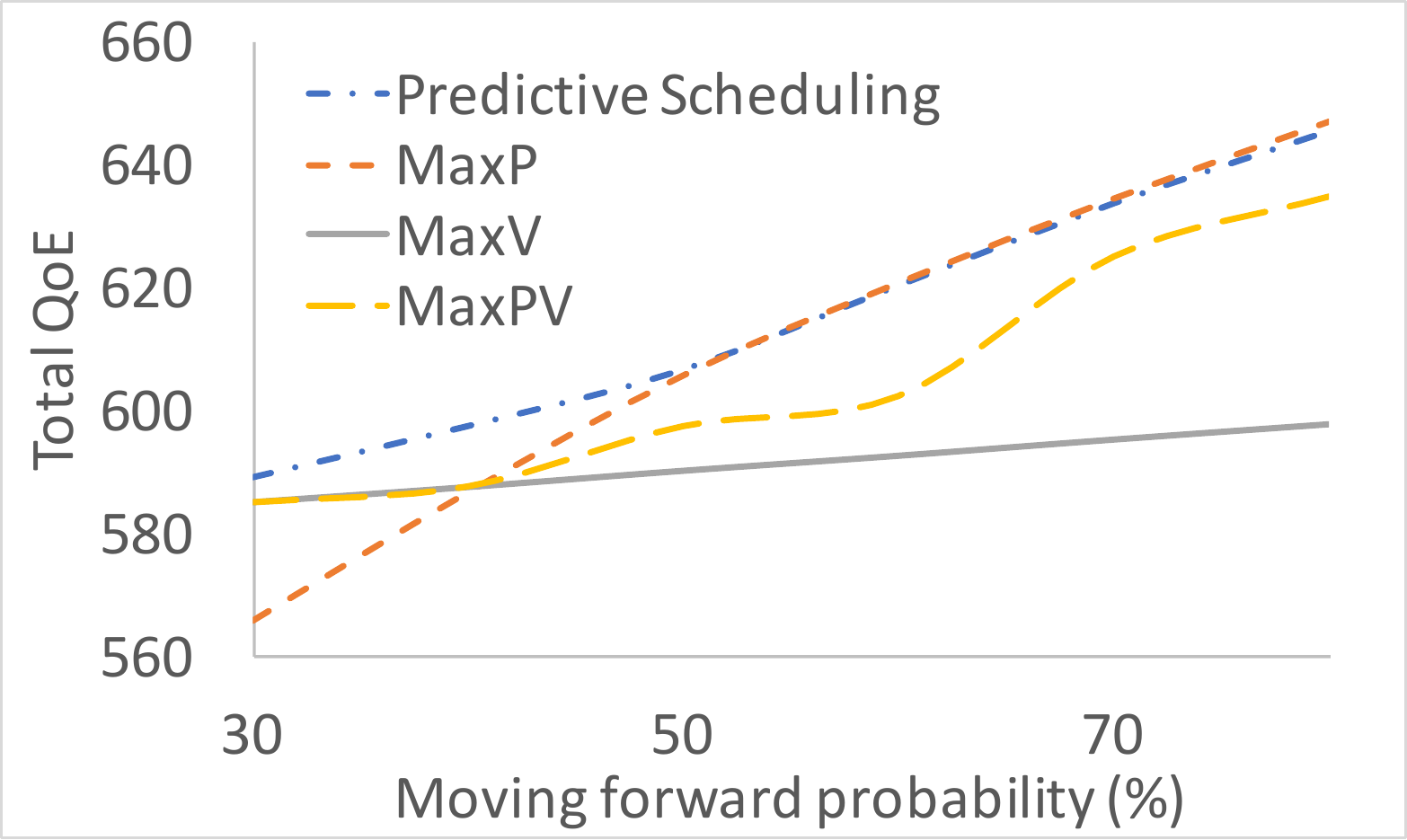}}
\hspace{0.01\linewidth} 
\subfigure[20 clients]{
\includegraphics[width=2.1in]{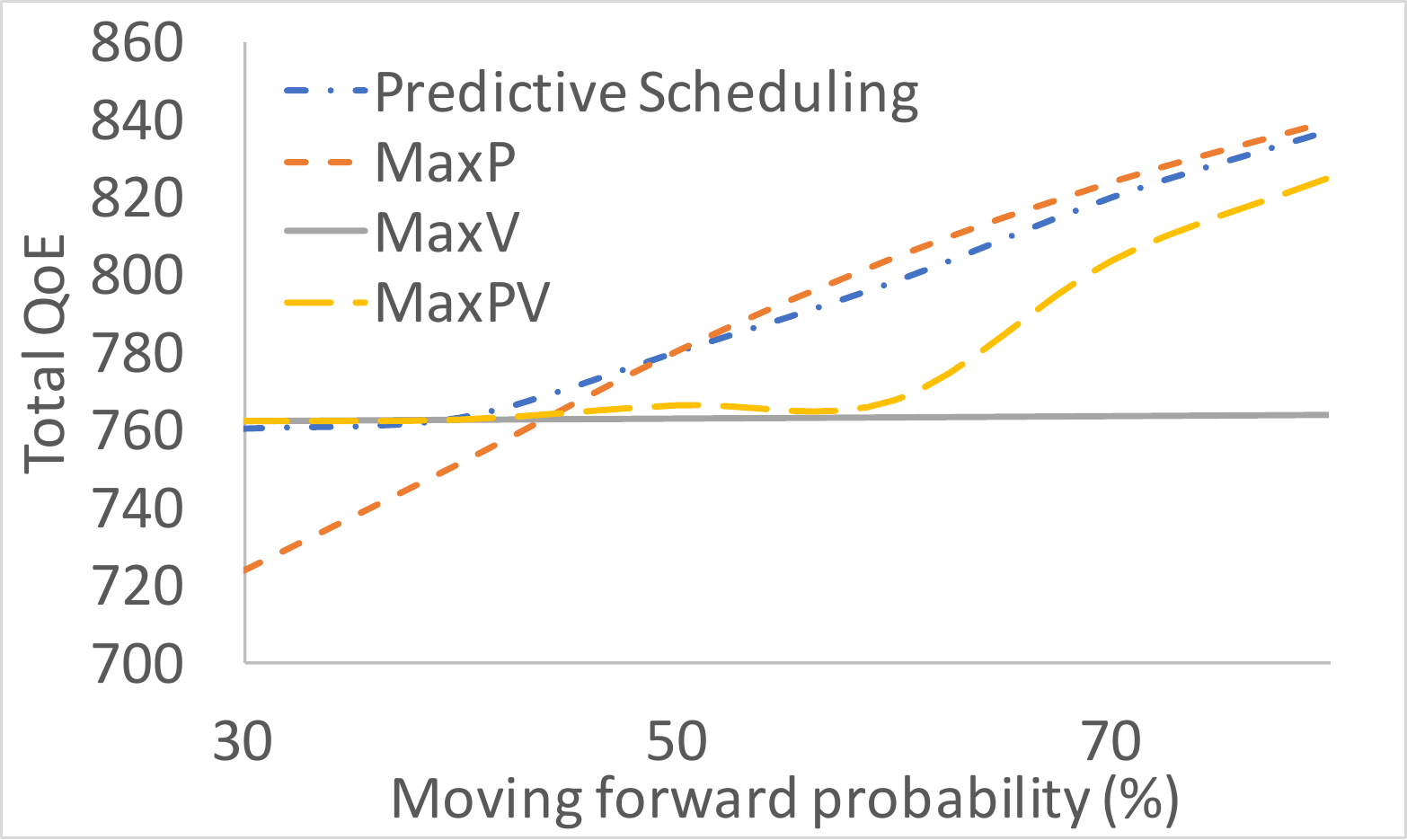}}
\caption{QoE comparison with $N_1=360$}\label{fig:simulation:N_1_360}
\end{figure*}

\begin{figure*}[t]
\subfigure[10 clients]{
\includegraphics[width=2.1in]{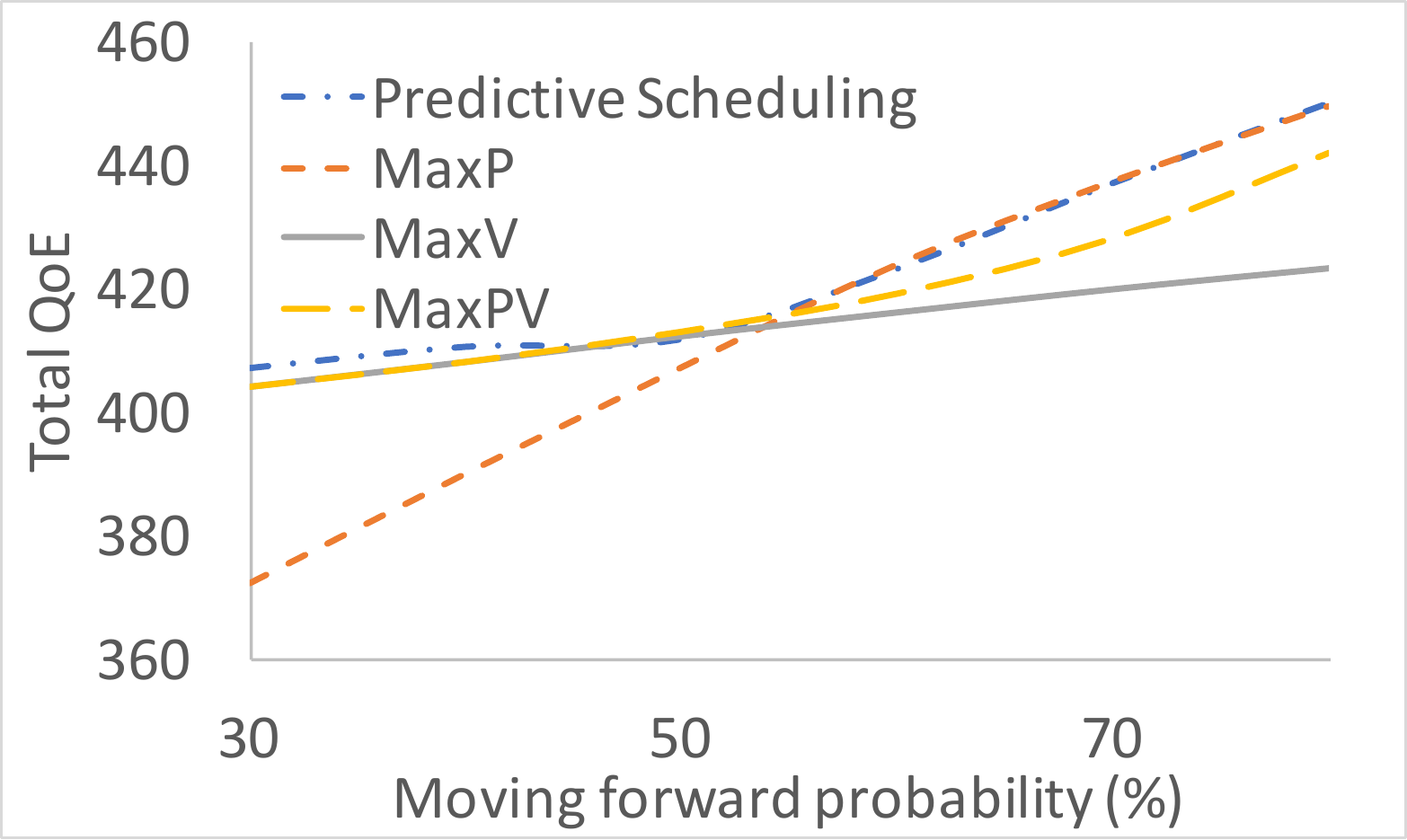}}
\hspace{0.01\linewidth} 
\subfigure[15 clients]{
\includegraphics[width=2.1in]{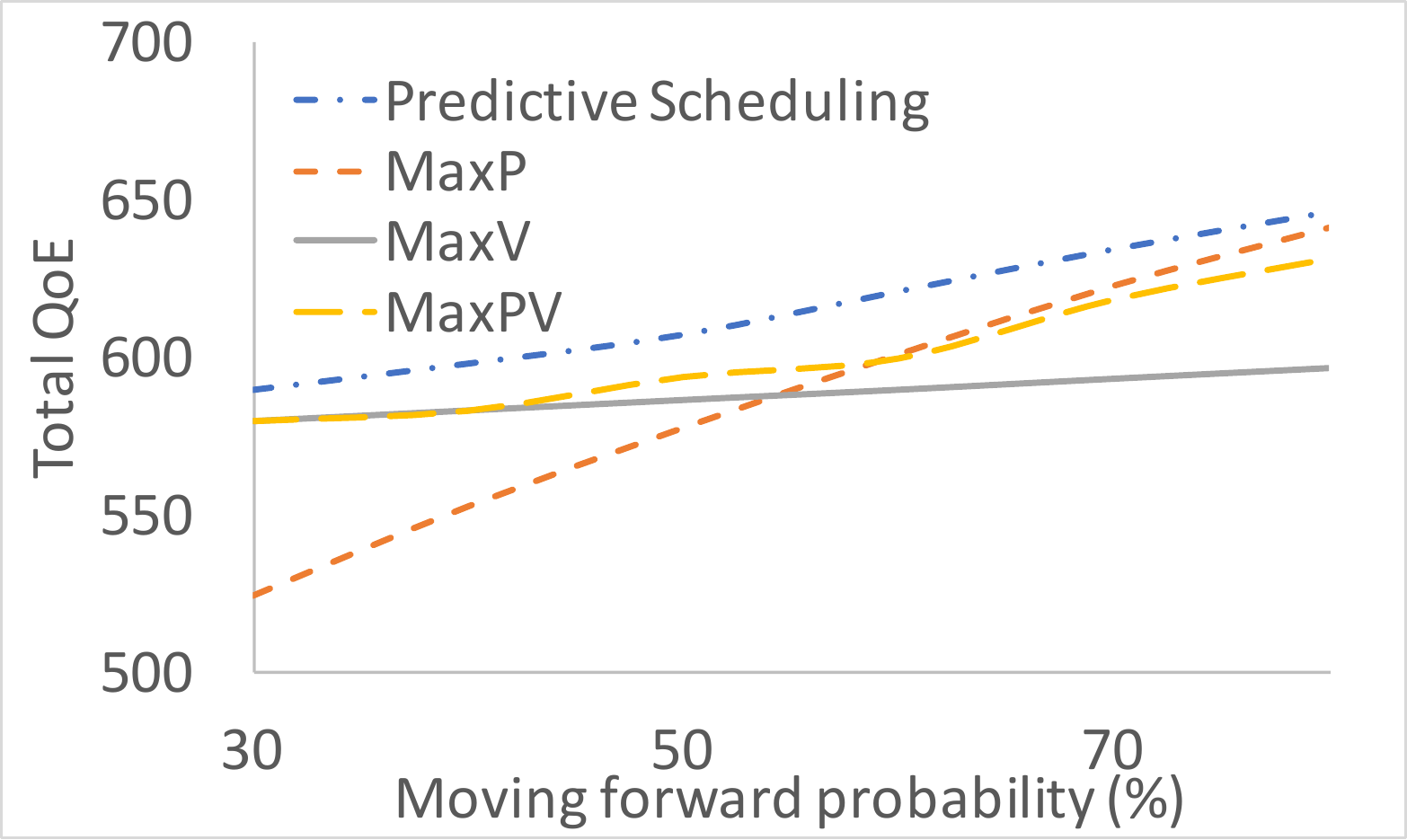}}
\hspace{0.01\linewidth} 
\subfigure[20 clients]{
\includegraphics[width=2.1in]{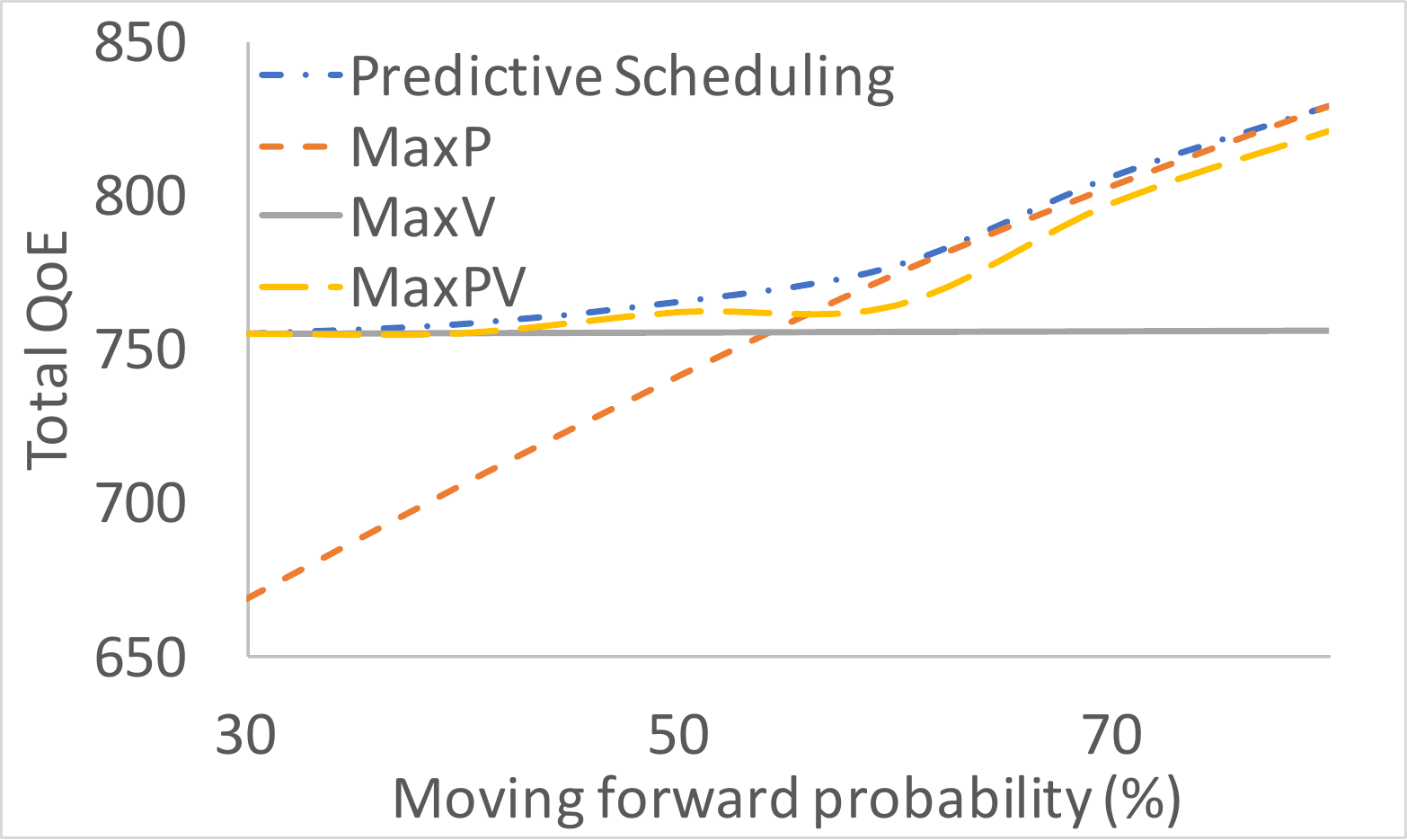}}
\caption{QoE comparison with $N_1=380$}\label{fig:simulation:N_1_380}
\end{figure*}

\textbf{VR environment and user mobility:} We model the VR environment as a two-dimensional space. Each user moves among the grid points in the space independently. In each interval, a user can only move to one of the four adjacent grid points. We consider a mobility model where a user's movement in an interval only depends on its current location and its movement in the previous interval. The user can either \emph{move forward}, which means its movement in the interval is in the same direction as that in the previous interval, \emph{turn left}, \emph{turn right}, or \emph{move backward}. We assume that the probabilities of the above four movements are $q$, $\frac{2(1-q)}{5}$, $\frac{2(1-q)}{5}$, and , $\frac{1-q}{5}$, respectively, where $q\in [0,1]$ is a simulation parameter.

\textbf{Evaluated policies:} In addition to our own policy, we have also implemented three other policies: The first is the \emph{MaxPV} policy, which transmits the $N_1$ packets with the highest expected contribution to QoE, i.e., $p_{s,i}v_i$, during the proactive scheduling phase. The second is the \emph{MaxP} policy, which transmits the $N_1$ packets with the highest $p_{s,i}$, and breaks ties by $v_i$, during the proactive scheduling phase. The last is the \emph{MaxV} policy, which transmits the $N_1$ packets with the highest $v_i$, and breaks ties by $p_{s,i}$, during the proactive scheduling phase. All three policies use the same policy as ours during the deadline scheduling policy, since it is obviously optimal.

\subsection{Performance Evaluation}

We now present our simulation results. All results presented in this section are the average of 100 simulation runs.

Fig. \ref{fig:simulation:N_1_360} and \ref{fig:simulation:N_1_380} demonstrate the performance comparison between the four evaluated policies under different settings, where the value of $N_1$ can be either 360 or 380, the number of clients ranges from 10 to 20, and the value of $q$, i.e., the probability that a client moves forward in an interval, ranges from 30$\%$ to 80$\%$. The motion-to-photon latency can be as small as 1.6 ms when $N_1=360$, and as small as 0.8 ms when $N_1=380$. It can be shown that our policy achieves the best performance in all settings. Intuitively, as the MaxV policy makes scheduling decisions based mainly on $v_i$, and not on $p_{s,i}$, it performs better when the probabilities of the four movements are more uniform, or, equivalently, when the probability of moving forward is smaller. On the other hand, as the MaxP policy strongly favors the movement with the highest likelihood, it performs better when the probability of moving forward is closer to $100\%$. One can indeed see such behaviors in the simulation results. One can also observe that our policy has similar performance as MaxV when $q$ is small, and has similar performance as MaxP when $q$ is large.

\begin{figure*}[t]
\subfigure[10 clients]{
\includegraphics[width=2.1in]{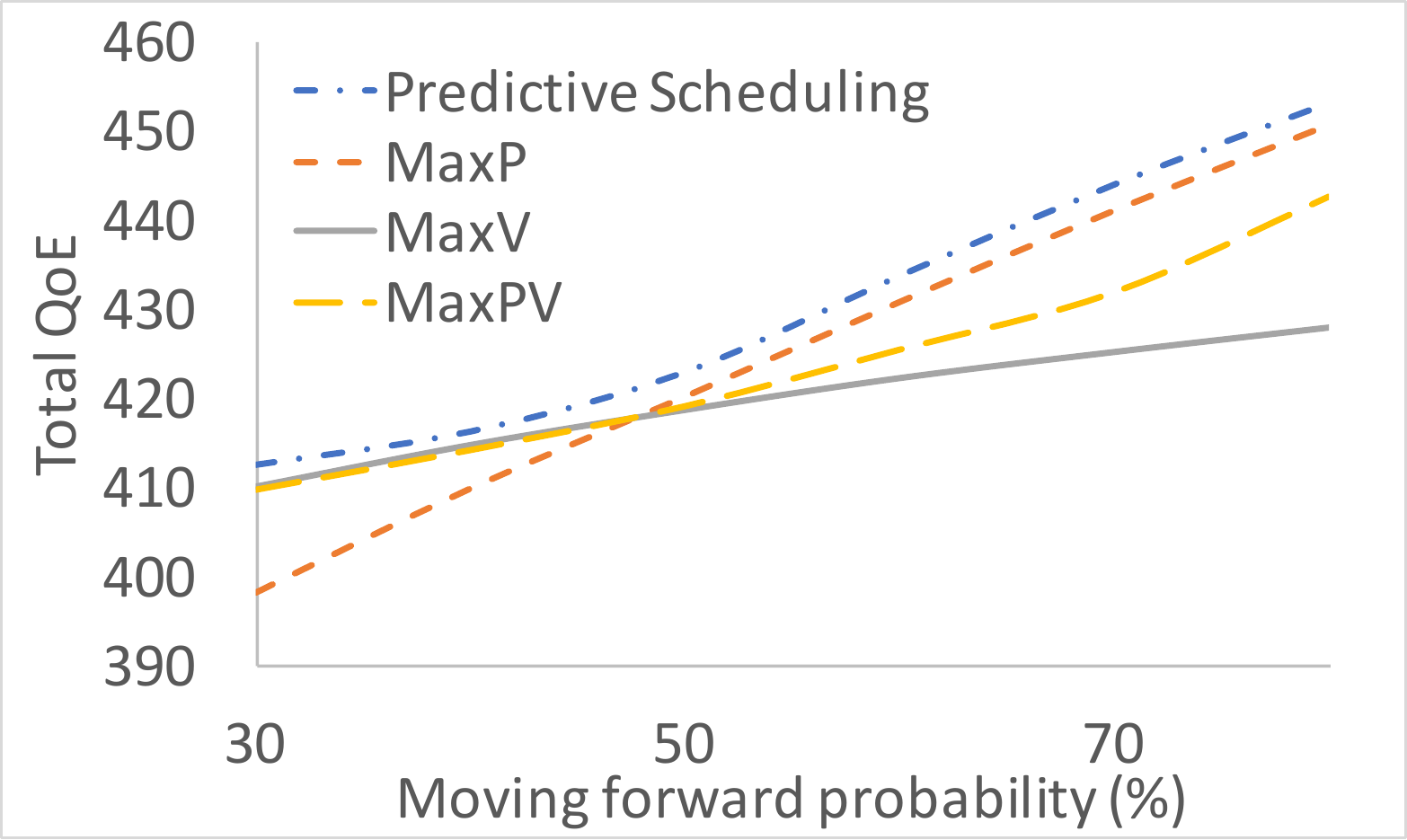}}
\hspace{0.01\linewidth} 
\subfigure[15 clients]{
\includegraphics[width=2.1in]{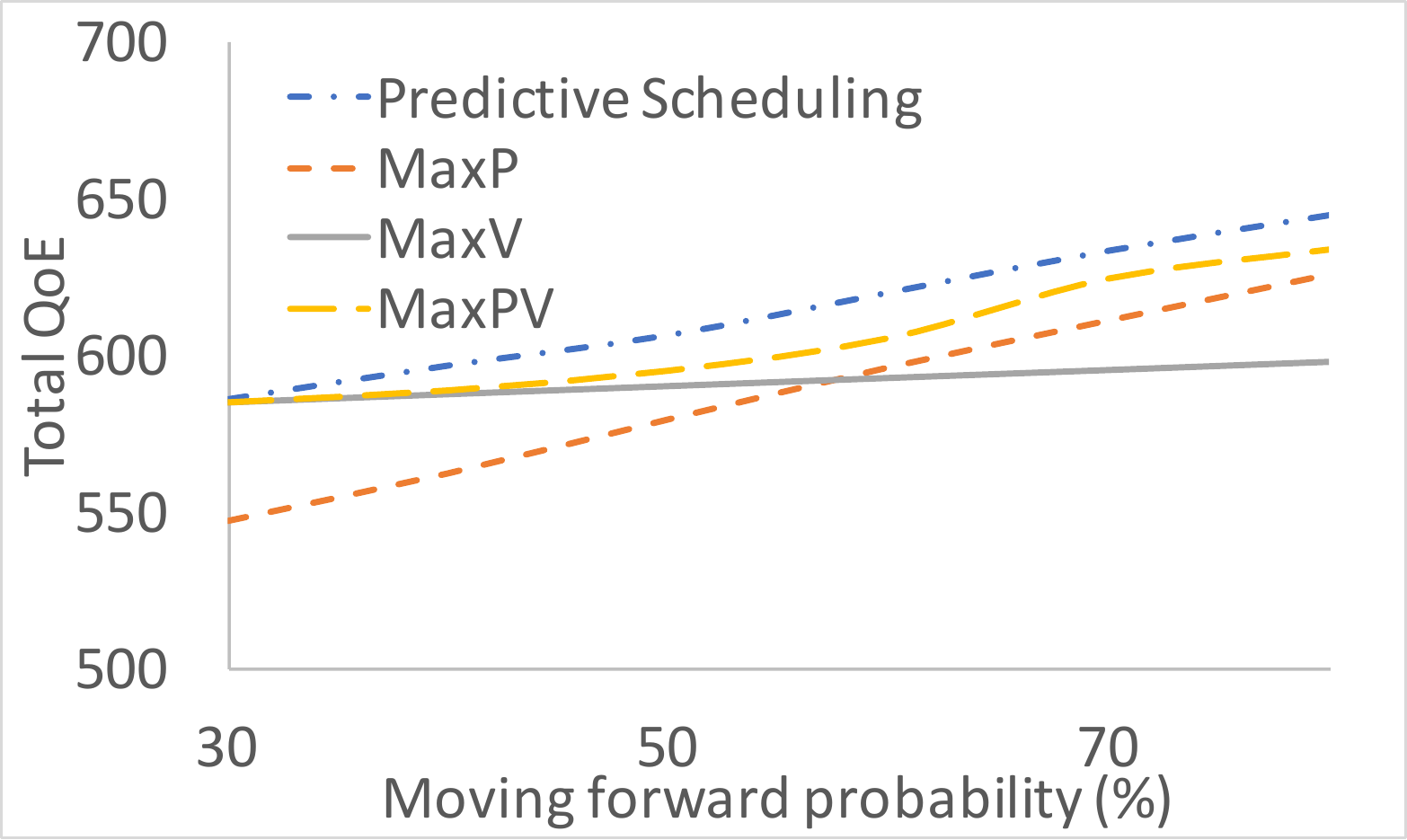}}
\hspace{0.01\linewidth} 
\subfigure[20 clients]{
\includegraphics[width=2.1in]{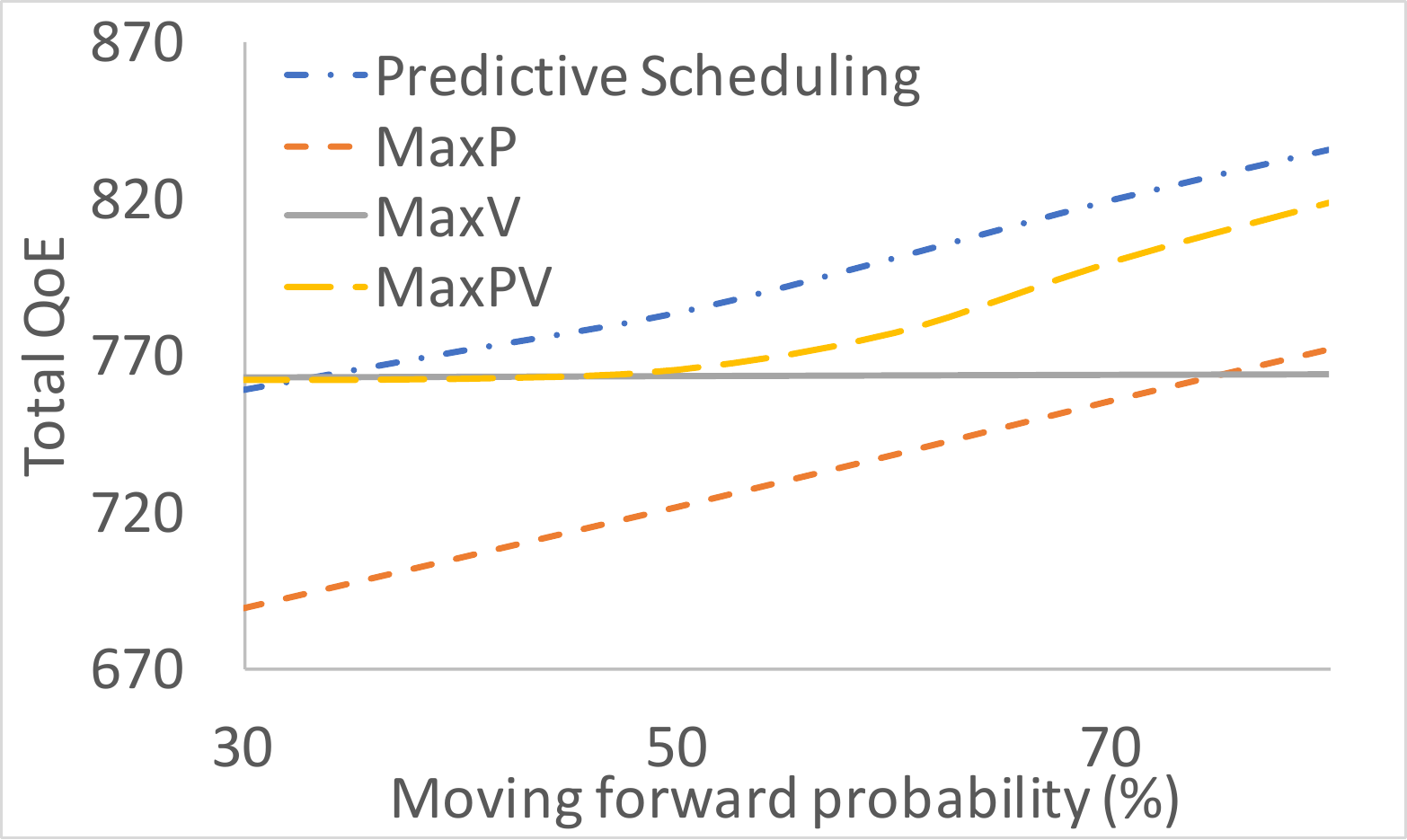}}
\caption{Simulation results with imperfect knowledge}\label{fig:simulation:witherror}
\end{figure*}

Next, we evaluate the impact of imperfect knowledge about $p_{s,i}$ to the performance of different scheduling policies. In this simulation, when the probability of moving forward of client $i$ is $q$, the server actually thinks the probability of moving forward is $q+e_i$, where $e_i$ is a uniform random variable in $[-0.1, +0.1]$. $N_1$ is 360, the number of clients ranges from 10 to 20, and $q$ ranges from $30\%$ to $80\%$. Simulation results are shown in Fig. \ref{fig:simulation:witherror}. One can see that our policy still achieves the best performance under all settings. A more interesting observation comes from comparing Fig. \ref{fig:simulation:witherror} and Fig. \ref{fig:simulation:N_1_360}, which has the same setting but assumes that the AP has the precise knowledge about $q$. It can be shown that the performance of our policy is virtually identical in these two figures. This suggests that our policy is very robust against some errors in estimating clients' mobility models. The reason that our policy is robust to such errors lies in the update of $\lambda$, i.e., line 25 in Alg. \ref{alg:predictive}. In our policy, $\lambda$ is updated based on $z_i(t)$, which only depends on the actual realization of client movements and does not involve $p_{s,i}$ at all. As such, an error in $p_{s,i}$ only has limited impact to our policy.

\begin{figure*}[t]
\subfigure[10 clients]{
\includegraphics[width=2.1in]{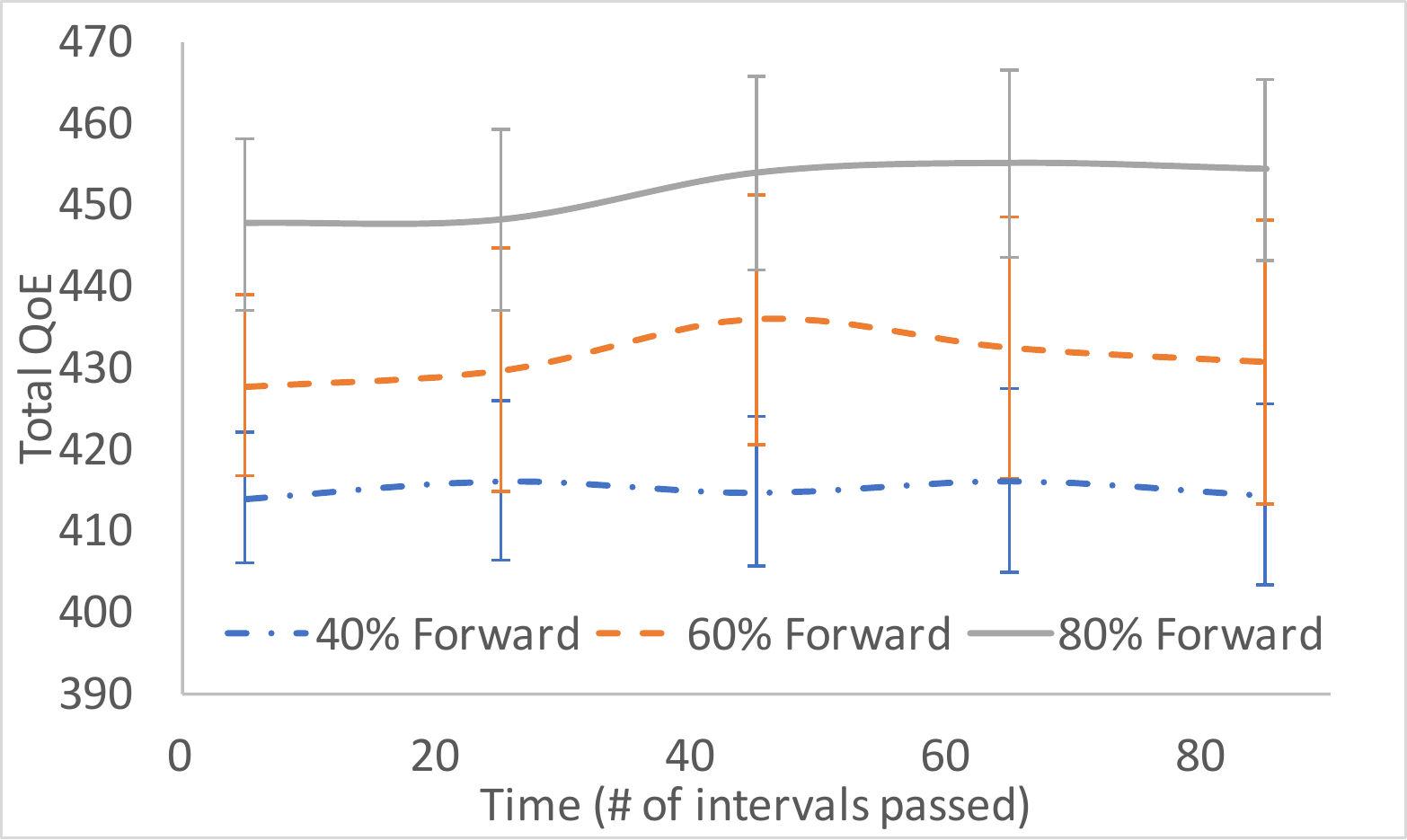}}
\hspace{0.01\linewidth} 
\subfigure[15 clients]{
\includegraphics[width=2.1in]{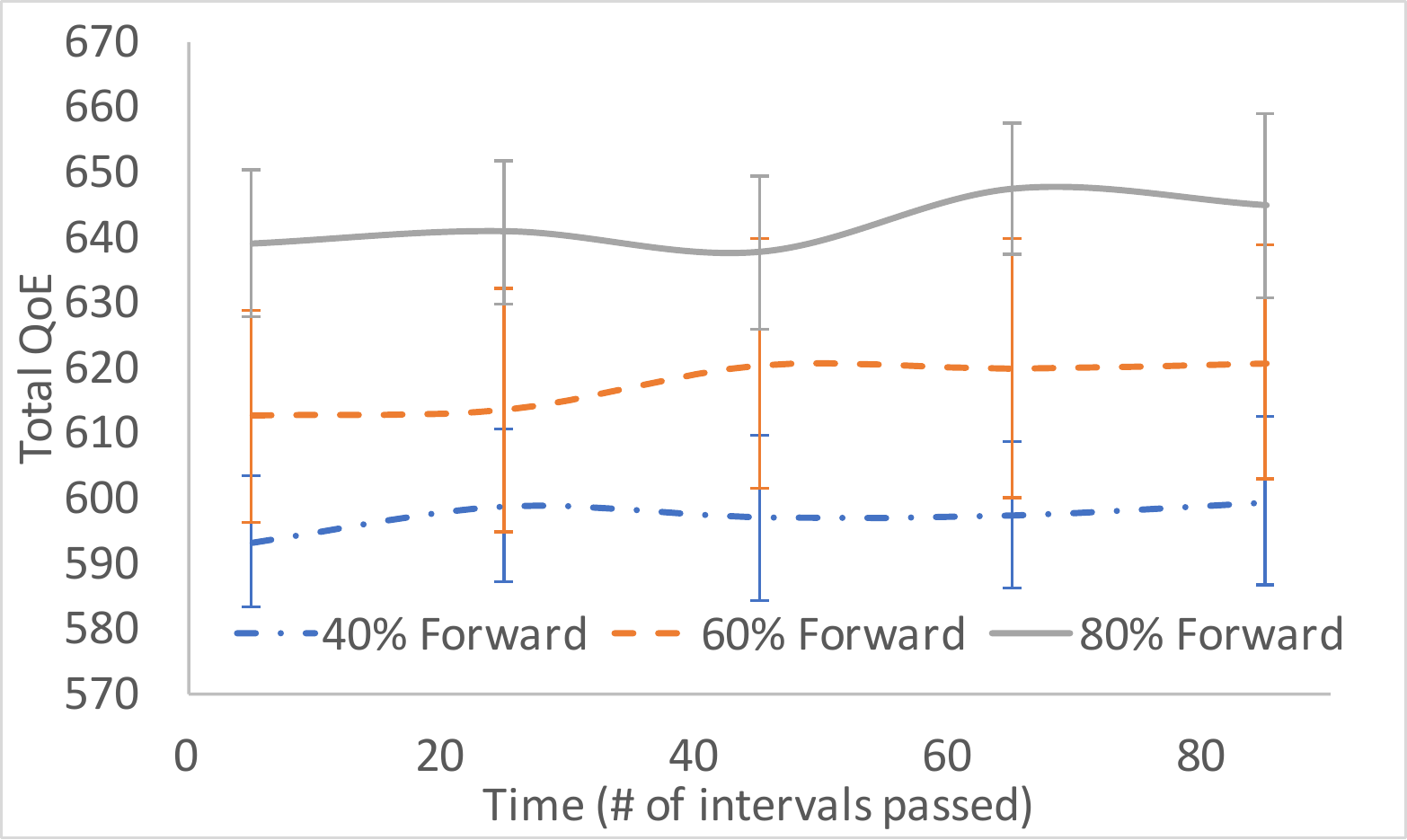}}
\hspace{0.01\linewidth} 
\subfigure[20 clients]{
\includegraphics[width=2.1in]{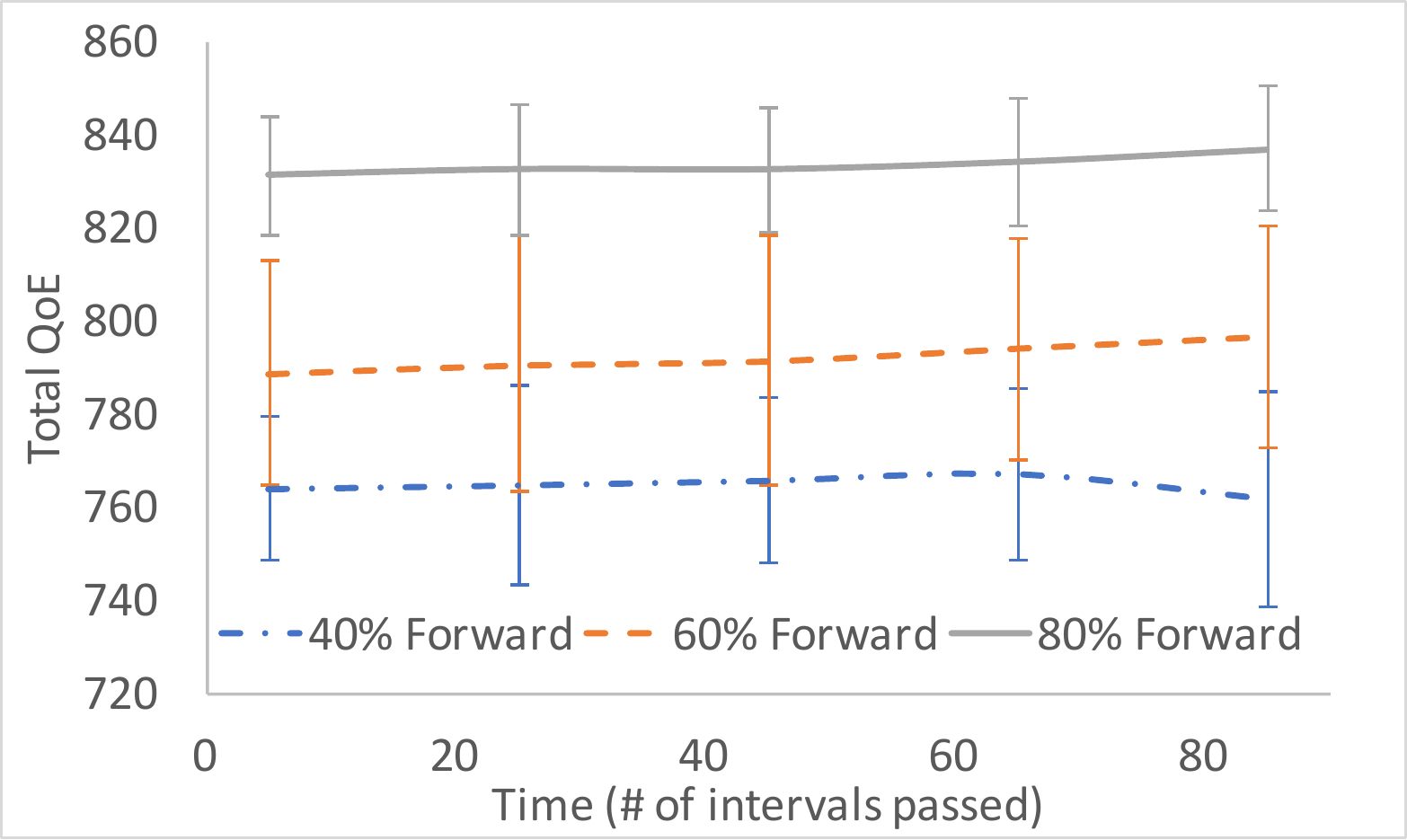}}
\caption{Instantaneous QoE over time}\label{fig:simulation:convergence}
\end{figure*}

Finally, we evaluate the convergence speed of our policy. Fig. \ref{fig:simulation:convergence} plots the average instantaneous QoE versus the number of intervals passed, with error bars indicating one standard deviation. One can see that the instantaneous QoE after merely five intervals, or, less than 0.1 second, is very close to that at steady state. This suggests that our policy converges very fast.
\section{Prototype Implementation} \label{section:prototype}

We have implemented a prototype of a VR system with predictive scheduling under Furion \cite{lai2019furion}, a Unity-based system that uses prefetching to enable untethered VR gaming on a smartphone with limited processing power. The default policy of Furion requires very high bandwidth to support high-quality VR gaming. Our implementation demonstrates that our proposed predictive scheduling policy allows for high-quality VR gaming even when the wireless bandwidth is limited. In this section, we first briefly describe the software architecture of Furion as well as its limitations. We then discuss adapting our proposed policy for Furion and demonstrate a side-by-side comparison between our policy and the default policy of Furion.

To achieve the required QoE (i.e., high responsiveness and high visual quality) for an untethered VR system, Furion divides the VR rendering workload into foreground interactions (FI) and background environment (BE). Furion employs a cooperative rendering module, where lightweight FI is rendered on mobile GPU and heavyweight BE is pre-rendered and pre-encoded on the server. In offline preprocessing, Unity rendering engine on server first pre-renders panoramic frames for the all-possible reachable area in BE (e.g., app-specific parameter). Then all these panoramic frames are pre-encoded into individual I frames using x264 \cite{x264} library (with constant rate factor 28 and fastdecode tuning enabled). During game play, depending on the player movement, Furion client running on a commodity phone, asks the Furion server for the next neighbouring frames and prefetches the I-frames for those neighbouring gridpoints using a persistent TCP connection. Meanwhile, Furion client parallely decodes the pre-fetched compressed BE frames using GStreamer Media SDK \cite{GStreamer} (with GPU acceleration enabled) and puts the decoded frames on a $360^{o}$ spherical movie texture around the rendered FI.

An important constraint of Furion is that its video decoder, GStreamer Media SDK \cite{GStreamer}, does not support multi-layered video coding. Instead, we assume that the panoramic frame for each location is encoded into three different resolutions: 512X256, 1024X512, 2048X1024 pixels (512p, 1024p, and 2K, respectively). The average file size is listed in Table~\ref{tab:QoE-resolution}. As shown in the table, we set the QoE of the three resolutions to be 512, 1024, and 2048, respectively. One can see that, while a high resolution frame has a high QoE, the ratio between QoE and file size is small. We also introduce a \emph{null} resolution with size 0 and QoE 0. The QoE of a client in each interval is then the QoE of the highest resolution frame that the client has received.

\begin{table}[h!]
\centering
\begin{tabular}{l|r|r|r}
Resolution & File-size (KB) & QoE & QoE-per-KByte    \\ \hline\hline
512p & 15.07 & 512 & 33.97\\ \hline
1024p  & 50.18 & 1024 & 20.4\\ \hline
2K  & 185.86 & 2048 & 11.02
\end{tabular}
\caption{Parameters for different resolutions} \label{tab:QoE-resolution}
\end{table}

\begin{figure*}[t]
\subfigure[Predictive Scheduling Policy]{
\includegraphics[width=3.3in]{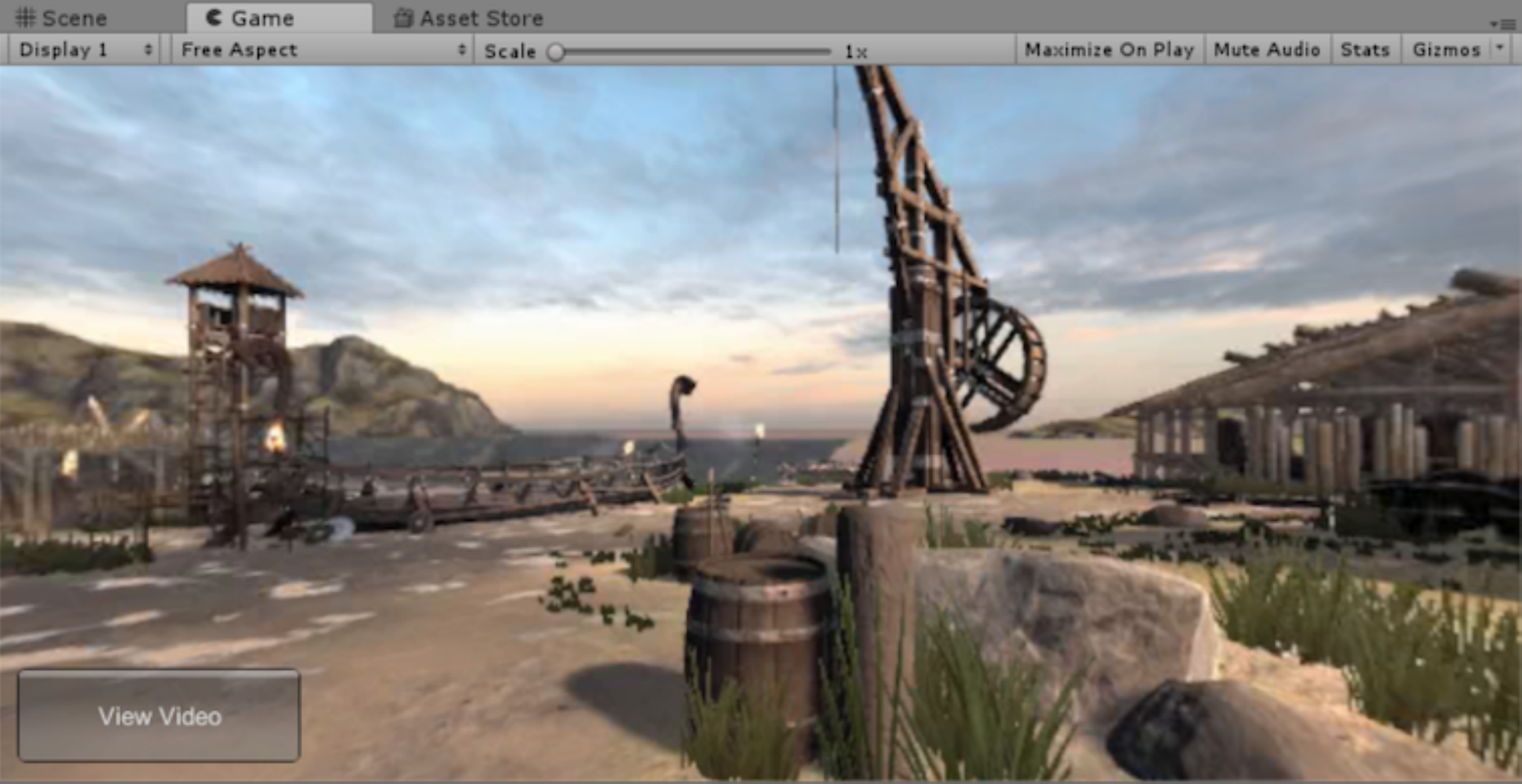}}
\hspace{0.01\linewidth} 
\subfigure[Default Furion Policy]{
\includegraphics[width=3.3in]{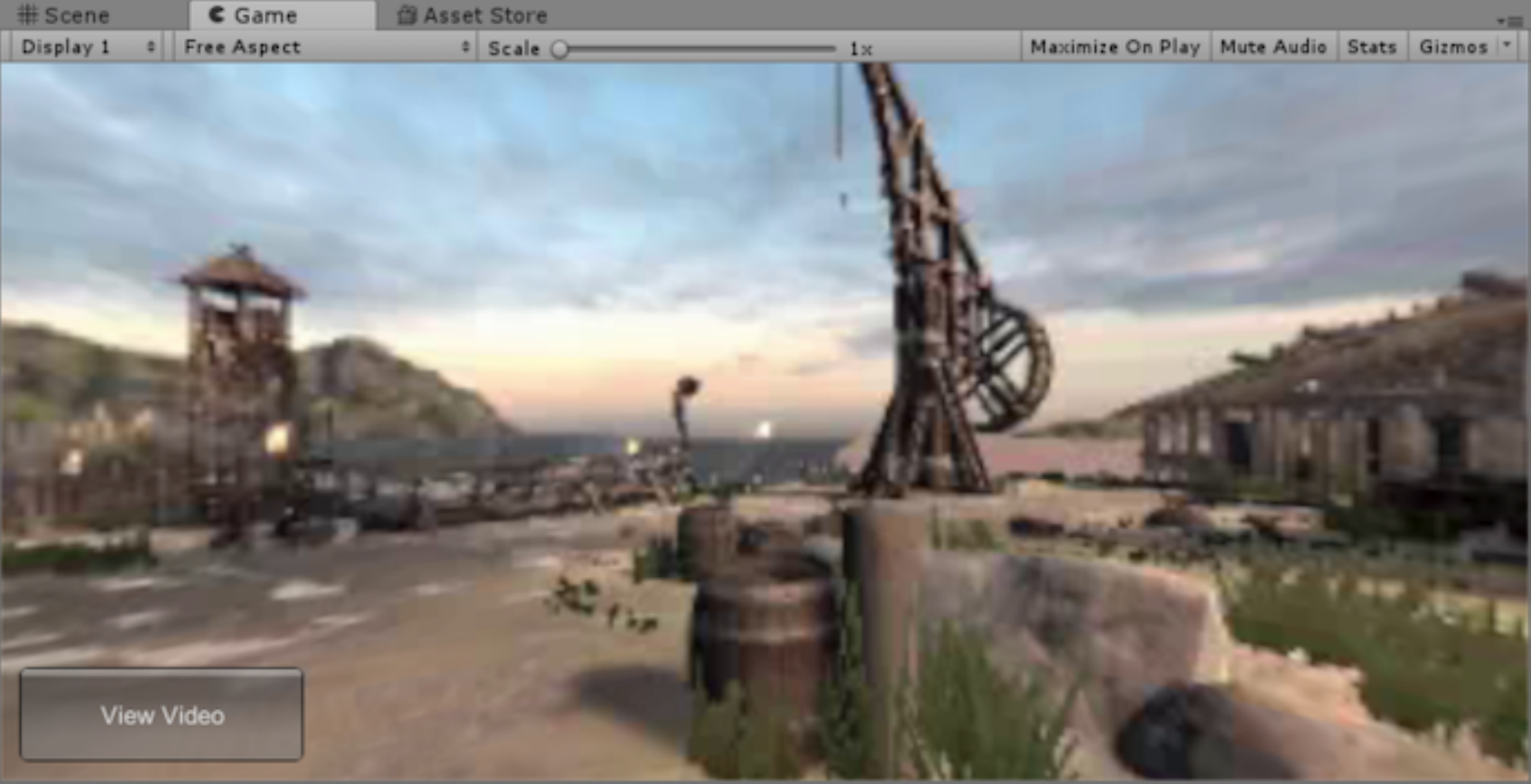}}
\caption{Screenshots of the prototype}\label{fig:screenshot}
\end{figure*}

Another challenge is that, as an application, Furion has no control over the underlying wireless protocol. Instead, we assume that the underlying wireless protocol transmits data in a first-in-first-out fashion, which is the default mechanism of virtually all existing wireless devices. We also assume that the Furion server can estimate network capacity based on long-term statistics. Specifically, we assume that the AP can transmit $N_1$ KBytes (instead of packets) in the proactive scheduling phase, and $N_2$ KBytes in the deadline scheduling phase. In each phase, the Furion server will select several frames and forward them to the underlying wireless protocol for transmission, with the constraint that the total size of the selected frames is no larger than $N_1$, and $N_2$, in the proactive scheduling phase, and the deadline scheduling phase, respectively.

We now discuss how to adapt our predictive scheduling policy for Furion. Let $\theta_i$ be the size of a frame $i$ with a specific resolution, and let $v_i$ be the QoE of frame $i$. Therefore, the \emph{QoE-per-KByte} of frame $i$ is $\frac{v_i}{\theta_i}$. In the same spirit of (\ref{19}), we define $\alpha_{s,i}$ by

\begin{equation} \label{equation:prototype:alpha}
\begin{split}
\alpha_{s,i} = 
    \begin{cases}
      \lambda \theta_i p_{s,i}, \quad \; \text{if} \quad \frac{v_i}{\theta_i} \geq \lambda,\\
       v_i p_{s,i}, \quad \text{if} \quad \frac{v_i}{\theta_i}<\lambda.
    \end{cases} \\
\end{split}
\end{equation}

In the proactive scheduling phase, the Furion server simply chooses a resolution for each adjacent location such that the sum of $\alpha_{s,i}$ over chosen frames is maximized, under the constraint the the sum of $\theta_i$ over chosen frames is no larger than $N_1$. This is a simple knapsack problem that can be solved efficiently. On the other hand, in the deadline scheduling phase, the Furion server transmits frames that offer the most improvement of QoE over the frames that have already been delivered, under the constraint that it can at most transmit $N_2$ KBytes. This is yet another simple knapsack problem.

We have implemented this algorithm on a single player Furion system. This system consists of one Furion server with wired connection to a TP-LINK TL-WR1043ND WiFi router and one Furion client that communicates with the WiFi router over wireless. We configure the network so that the Furion server can transmit 200 KBytes in the proactive scheduling phase, and 60 KBytes in the deadline scheduling phase. As for the player's mobility pattern, we assume that the player has a 70\% chance of moving forward, and 10\% chance of moving in each of the remaining three directions. 

Fig. \ref{fig:screenshot} shows side-by-side the screenshots of our algorithm and that of the default Furion policy. It is clear that our algorithm displays a higher resolution frame. The default Furion policy does not take the mobility pattern of the player into account. As a result, it needs to evenly allocate the bandwidth among the four adjacent locations, and transmits the 1024p frame for each of them. On the other hand, knowing that the player has the highest probability of moving forward, our algorithm devotes all bandwidth in the proactive scheduling phase to transmit the 2K frame for the forward location. Even when the player makes a sudden turn, our algorithm is still able to transmit the medium-resolution frame in the deadline scheduling phase.

\section{Related Work}  \label{section:related}

Enabling untethered high-quality interactive VR on today's resource-constrained commodity devices has attracted strong interest both from academia and industry. MoVR \cite{abari2017enabling, abari2016cutting} proposes to cut the cord (i.e. tethered HDMI cable) by using high bandwidth mmWave technologies (e.g. 802.11ad/WiGig). Flashback  \cite{boos2016flashback} pre-renders the whole virtual environment and stores them on the local cache of the mobile device in order to reduce the rendering workload on commodity devices. However, Flashback doesn't support interactive VR applications and also requires overwhelming storage on mobile devices. 
Liu et al. \cite{liu2018cutting} adopt a thin-client framework employing VSync driven parallel rendering and streaming to achieve high-quality VR on mobile devices. This system might suffer from the frame misses due to adjustment error of the rendering time on the server according to the VSync information feedback from the client. In order to reduce the high network transmission size, a few other existing approaches \cite{bao2016shooting, hosseini2016adaptive, rondao201716k} employ transmitting only the pixel information inside the player's Field-of-View (FoV) through different tiling schemes. Different Tiling schemes based on HEVC is proven to be not the best fit in the latency perspective due to increased encoding and decoding time \cite{shi2019freedom}. Several recent works \cite{baig2019jigsaw, he2018rubiks, nasrabadi2017adaptive} attack this problem with the new layered video coding methods with encoding the visible area with higher bit-rate and others in lower bit rate. However, these approaches require player's FoV prediction which is more stringent than player's movement prediction due to an infinite number of possibilities of player's FoV and also can't guarantee good performance for all players in VR world.

Using prediction to pre-fetch contents has attracted growing research interests. Tadrous et al. \cite{tadrous2014proactive, tadrous2015optimal} use statistical prediction to pre-fetch contents during the off-peak hours with the goal of reducing bandwidth consumption during peak hours. Shoukry et al. \cite{shoukry2014proactive} develop a software architecture on smartphones that opportunistically pre-fetch contents when the smartphones are connected to WiFi to reduce bandwidth consumption over cellular networks. The prediction window of these studies is on the scale of hours, and hence these studies are not suitable for interactive VR, which operates in a very fast timescale. There are also some studies that aim to leverage prediction to improve the performance of real-time applications. Chen and Huang \cite{chen2018timely} study using prediction to increase the timely-throughput of real-time traffic. Yin et al. \cite{yin2019only} develop predictive scheduling policies to minimize age-of-information (AoI). Liu et al. \cite{liu2019proactive} analyze the delay performance of threshold-based predictive scheduling policies. As these studies do not explicitly consider the impact of multi-layered video coding, they cannot be directly applied to interactive VR applications, where different packets of the same flow can have significantly different impacts on QoE.
\section{Conclusion} \label{section:conclusion}

We have studied the problem of improving QoE for interactive VR applications through predictive scheduling. By studying an analytical model that jointly captures the characteristics of various VR components, we have developed an optimal predictive scheduling policy. Simulation results show that our policy significantly outperforms others. It is also resilient to prediction errors and converges very fast. Furthermore, we build a prototype system to test our policy. Experimental results show that our policy is able to deliver much better video quality.

\section{Acknowledgement}
This material is based upon work supported in part by NSF and Intel under contract numbers CNS-1719384, CNS-1719369, CNS-1719371, in part by NSF under contract number CNS-1824337, in part by the U.S. Army Research Laboratory and the U.S. Army Research Office under contract/grant number W911NF-18-1-0331, and in part by Office of Naval Research under contracts N00014-18-1-2048 and N00014-19-1-2621.

\balance
\bibliographystyle{ieeetr}
\bibliography{sigproc} 

\end{document}